\documentclass[reprint,superscriptaddress,amsmath,amssymb,prb,nolongbibliography]{revtex4-2}
\usepackage{bbold}
\usepackage{graphicx}
\usepackage{dcolumn}
\usepackage{bm}
\usepackage[svgnames]{xcolor}
\usepackage[colorlinks=True,linkcolor=DarkRed,citecolor=ForestGreen,urlcolor=MediumBlue,pdfstartview=FitH,bookmarks=False,pdfpagemode=UseNone]{hyperref}
\usepackage{physics}
\usepackage[normalem]{ulem}

\begin{document}

\title{\boldmath Can $dd$ excitations mediate pairing ?}
\homepage[{Preprint of }]{Physica C 613, 1354321 (2023),  Open Access,  https://doi.org/10.1016/j.physc.2023.1354321}

\author{Francesco Barantani}
    \affiliation{Department of Quantum Matter Physics, University of Geneva, 1211 Geneva, Switzerland}
    \affiliation{Institute of Physics, \'Ecole Polytechnique F\'ed\'erale de Lausanne, Lausanne, 1015, Switzerland}
\author{Christophe Berthod}
    \affiliation{Department of Quantum Matter Physics, University of Geneva, 1211 Geneva, Switzerland}
\author{Dirk van der Marel}
 \email{dirk.vandermarel@unige.ch} 
    \affiliation{Department of Quantum Matter Physics, University of Geneva, 1211 Geneva, Switzerland} 

\date{\today}

\begin{abstract}
The Cu-$3d$ states in the high-$T_c$ cuprates are often described as a single band of $3d_{x^2-y^2}$ states, with the other four $3d$ states having about 2 to 3 eV higher energy due to the lower-than-octahedral crystal field at the copper sites. However, excitations to these higher energy states observed with RIXS show indications of strong coupling to doped holes in the $3d_{x^2-y^2}$ band. This relaunches a decades-old question of the possible role of the orbital degrees of freedom that once motivated Bednorz and M\"uller to search for superconductivity in these systems. Here we explore a direction different from the Jahn-Teller electron-phonon coupling considered by Bednorz and M\"uller, namely the interaction between holes mediated by $dd$ excitations.  
\end{abstract}

\maketitle

\section{Introduction}

In 1986 it was widely believed that superconductivity above the then-known limit of 23 Kelvin was not possible. Who could have predicted that, initiated by Georg Bednorz and Karl Alex M\"uller's discovery, superconductivity in the cuprates would go through the liquid nitrogen ceiling. The consequences for fundamental science and for applications are numerous and will be detailed throughout the tributes in this special issue.

\begin{figure}[tb]
    \centering
    \includegraphics[width=\columnwidth]{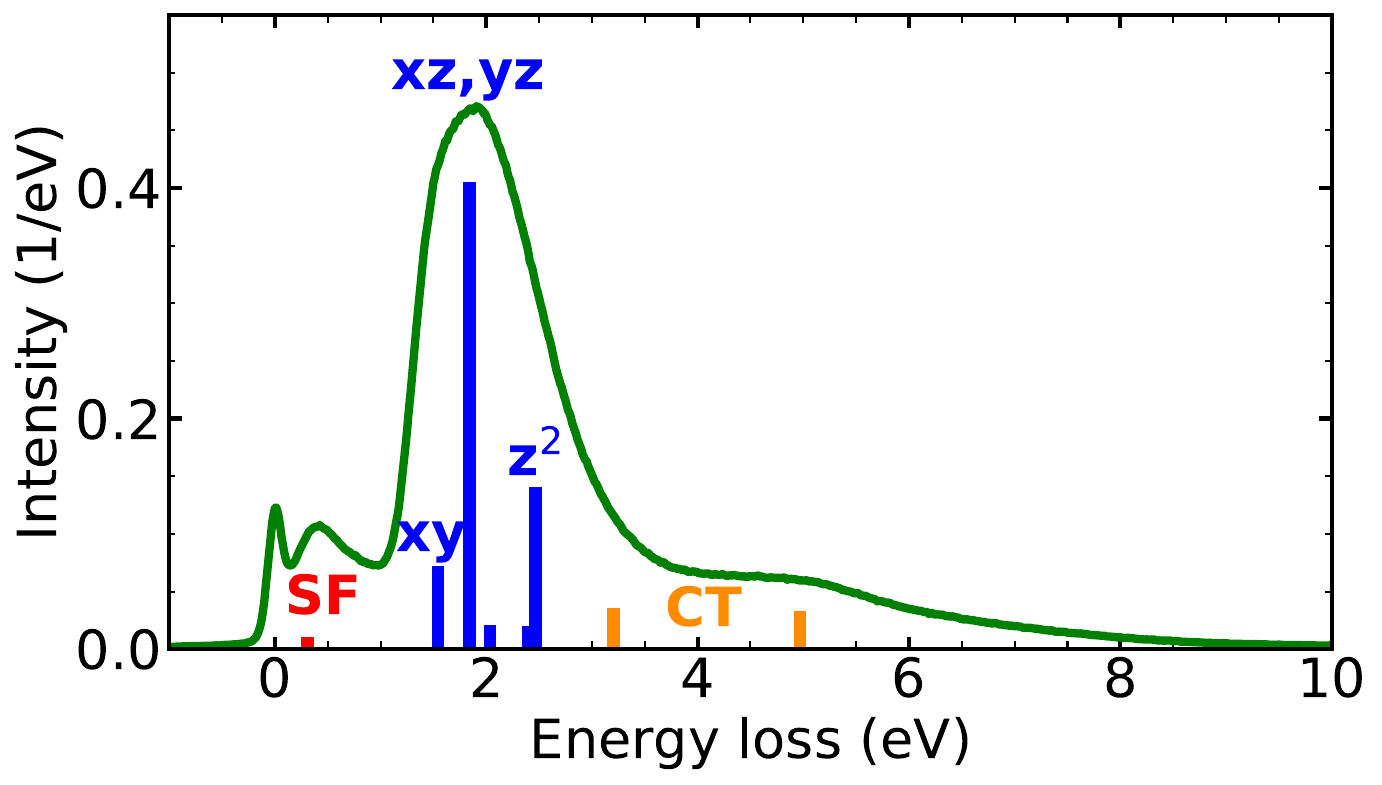}
    \caption{RIXS spectrum of overdoped Bi$_2$Sr$_2$CaCu$_2$O$_{8-x}$ (Bi-2212) ($T_c=70$~K). Labels refer to spin (red), $dd$ (blue) and charge transfer (orange) excitations. See Ref.~\cite{barantani2022} for details. }
    \label{fig:rixs}
\end{figure}

In their Nobel lecture \cite{bednorz1987} Bednorz and M\"uller provided a captivating account of the path that led them to their discovery. A remarkable aspect was the consideration of the 3$d$ orbital degrees of freedom:  ``For Cu$^{3+}$ with 3$d^8$ configuration, the orbitals transforming as base functions of the cubic $e_g$ group are half-filled, thus a singlet ground state is formed. In the presence of Cu$^{2+}$ with 3$d^9$ configuration the ground state is degenerate, and a spontaneous distortion of the octahedron occurs to remove this degeneracy. This is known as the Jahn-Teller effect.''  For a Cu ion surrounded by O$^{2-}$ ligands with an octahedral crystal field, the 3$d$-levels are grouped in $\ket{\underline{e}_g}$ ({\em i.e.} $\ket{\underline{d}_{x^2-y^2}}$, $\ket{\underline{d}_{3r^2-z^2}}$) and $\ket{\underline{t}_{2g}}$ ({\em i.e.} $\ket{\underline{d}_{xy}}$, $\ket{\underline{d}_{yz}}$, $\ket{\underline{d}_{zx}}$) levels at about 2~eV higher energy. Here the notation $\ket{\underline{d}_{j}}$ refers to a hole in an otherwise fully occupied 3$d$-shell. In the cuprate superconductors the crystal field  {symmetry} at the Cu sites is lower than octahedral. This causes the ground state to become $\ket{\underline{d}_{x^2-y^2}}$, with $\ket{\underline{d}_{3r^2-z^2}}$ at least 1~eV above the ground state. This splitting can be directly observed with resonant inelastic x-ray scattering (RIXS). An example is provided in Fig.~\ref{fig:rixs} \cite{barantani2022} for the hole-doped Bi-based cuprate. Since the Jahn-Teller effect relies on the two states being degenerate, it is tempting to conclude that this effect does not play a role here. However, the Jahn-Teller effect is restored --- at least in part --- if the $d_{3r^2-z^2}$ and $d_{x^2-y^2}$ bands are coupled (citing K.~A.~M\"uller \cite{muller2007}) ``via vibronic interactions, {\em i.e.} the ions are allowed to move \cite{bussmann2005}. The latter coupling lowers the upper band to within a few meV of the lower one.'' In the RIXS spectrum this should be revealed by a strong broadening of the $\ket{\underline{d}_{3r^2-z^2}}$ state, so that the zero-phonon line occurs close to zero RIXS excitation energy. This effect was observed by Lee \textit{et al.}\ for the case of Ca$_2$Y$_2$Cu$_5$O$_{10}$ \cite{lee2014}. However, they located the zero-phonon line only 0.2 eV below the peak maximum, indicating that the vibrational coupling is too weak to overcome the splitting between $\ket{\underline{d}_{3r^2-z^2}}$ and $\ket{\underline{d}_{x^2-y^2}}$. 

Recently the detailed temperature dependence of the $dd$ excitations in Bi-2212 has been studied by, among others, the present authors \cite{barantani2022} and it was found that the energies of these modes are influenced by the system becoming superconducting. Superconductivity-induced shifts of $+30$~meV were observed on the overdoped side, about $-12$~meV on the underdoped side, and no shift within the error bar for optimal doping. It was argued that these shifts --- including the sign-change as a function of doping --- could be caused by differences in local spin-correlations of the different phases. These observations also clearly indicate that there is a non-negligible coupling between the conducting holes in the cuprates and the $dd$ excitations.  

A $dd$ excitation is essentially a local orbital flip from the $\ket{\underline{d}_{x^2-y^2}}$ ground state to one of the other $\ket{\underline{d}_{j}}$ states ($j=xy,yz,zx$ or $3r^2-z^2$). From this perspective it is unavoidable that a coupling exists between the holes in the ${d}_{x^2-y^2}$ band and the $dd$ excitations, and that this coupling is in fact very strong. We interrupt the flow of logic to point out a commonality with Bednorz and M\"uller's intuition, namely that the 3$d$ orbital degrees of freedom could be one of the key ingredients. Instead of coupling the $d$ orbital degrees of freedom to vibrations, we will now consider the coupling between $dd$ excitations and conduction electrons or holes. The question then is whether such coupling contributes to the pairing interaction, and whether this contribution is positive or negative. 

The possibility that $dd$ excitations in the cuprates could mediate superconducting pairing has been pioneered by {Werner} Weber and collaborators \cite{weber1988a,weber1988b} and by William Little and collaborators \cite{holcomb1994,holcomb1996,little2000,little2007}. Weber considered a model where conduction is confined to holes in the O-$2p$ band, where the copper ions are purely $3d^9$, and pairing is mediated by $dd$ orbital excitations within the $e_g$ manifold of the Cu $3d^9$. Little \textit{et al.}\ used the Eliashberg equations to describe the coupling of conduction electrons or holes to $dd$ excitations.  The approach that we explore here differs from the previous ones. Compared to the approach of Little \textit{et al.}\ we use a more direct description of the coupling to the $3d$ orbital degrees of freedom; and instead of Weber's model with separate O-$2p$ and Cu-$3d_{x^2-y^2}$ bands, we adopt the model of a single partly filled band of mixed Cu-$3d_{x^2-y^2}$--O-$2p$ character.

\section{The model}

\begin{figure}[b]
    \centering
    \includegraphics[width = \columnwidth]{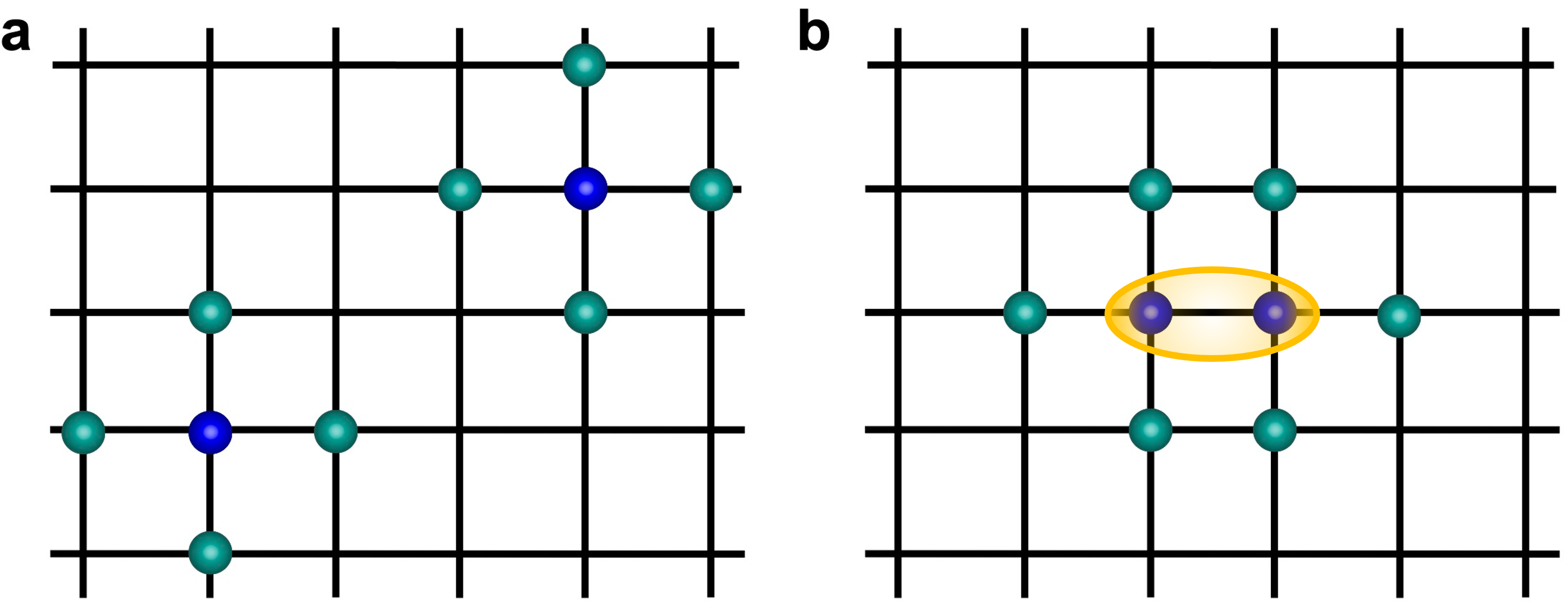}
    \caption{\textbf{a} Two additional holes present on distant sites (blue spheres) are renormalized by the interaction with 4 nearest neighbors each (green spheres). \textbf{b} Two additional holes present on neighboring sites are renormalized by the interaction with 6 nearest neighbors {\em and} by virtual hopping of particles between the two sites (yellow).}
    \label{fig:latticeExcitons}
\end{figure}

The question that we want to address is whether the coupling to $dd$ excitations could provide an attractive short-range interaction between the free carriers. We model the problem by considering a 2-dimensional square lattice of atoms. We then dope two additional holes into this lattice, (i) in a configuration where the holes are at infinite distance from each other giving the energy $E_{\infty}$, (ii) where they occupy nearest-neighbor sites giving the energy $E_{nn}$. The energy difference $E_{nn}-E_{\infty}$ then constitutes the effective interaction energy of a doped hole pair at a nearest-neighbor distance.

In the interest of simplicity, we will limit the Hilbert space on each site to two of the $3d$ orbitals, which is also the minimal model to consider $dd$ excitations. In the hole language, $3d_{x^2-y^2}$ ($\ket{a}$) is the state with the lowest energy, and we represent the other four orbitals at 2--3~eV above the ground state by a single state $\ket{b}$. In this subspace the local Hamiltonian reads \footnote{A more general expression of this Hamiltonian in the SU(2) invariant form can be found in Ref.~\cite{georges2013}.} 
\begin{multline}
    H = \sum_{\sigma}\big(\varepsilon_a^{} a_\sigma^\dagger a_\sigma^{} + \varepsilon_b^{} b_\sigma^\dagger b_\sigma^{}\big)
    + U \big(a_{\uparrow}^\dagger a_{\uparrow}^{}a_{\downarrow}^\dagger a_{\downarrow}^{} + b_{\uparrow}^\dagger b_{\uparrow}^{}b_{\downarrow}^\dagger b_{\downarrow}^{}\big)\\
    + (U-2J)\big(a_{\uparrow}^\dagger a_{\uparrow}^{}b_{\downarrow}^\dagger b_{\downarrow}^{}+a_{\downarrow}^\dagger a_{\downarrow}^{}b_{\uparrow}^\dagger b_{\uparrow}^{}\big)\\ 
    + (U-3J)\big(a_{\uparrow}^\dagger a_{\uparrow}^{}b_{\uparrow}^\dagger b_{\uparrow}^{}+a_{\downarrow}^\dagger a_{\downarrow}^{}b_{\downarrow}^\dagger b_{\downarrow}^{}\big)\\
    - J\big(a_{\uparrow}^\dagger a_{\downarrow}^{}b_{\downarrow}^\dagger b_{\uparrow}^{}+a_{\downarrow}^\dagger a_{\uparrow}^{}b_{\uparrow}^\dagger b_{\downarrow}^{}
        + a_{\uparrow}^\dagger a_{\downarrow}^\dagger b_{\uparrow}^{}b_{\downarrow}^{} + a_{\downarrow}^{} a_{\uparrow}^{}b_{\downarrow}^\dagger b_{\uparrow}^\dagger\big),
    \label{eq:2level-atom}
\end{multline}
where $a^{\dagger}$ ($a$) and $b^{\dagger}$ ($b$) are the creation (annihilation) operators for a hole with energy $\varepsilon_a$ and $\varepsilon_b$, respectively, $U = A+ 4B +3C$, $J = 5B/2+2C$ and $A, B, C$ are the Racah parameters. The $dd$ excitation corresponds to an excitation from $\ket{a}$ to $\ket{b}$, with excitation energy 
\begin{equation}
    \Omega_{ab}=\varepsilon_b-\varepsilon_a.
\end{equation}
The first term of Eq.~(\ref{eq:2level-atom}) is the kinetic energy, the second term proportional to $U$ is the intra-orbital Coulomb interaction, the third and fourth terms represent inter-orbital Coulomb and Hund's coupling, and the last term involves on-site inter-orbital spin-flips and on-site pair-wise orbital flips \cite{kanamori1963}.

A hole doped into the Mott-insulating state has the possibility of making virtual excursions to the neighboring sites, which can be either the same orbital $a$ or the orbital $b$ at a higher energy. When we dope two holes and keep them at a sufficiently long distance from each other, there is no interference between their respective virtual excursions, but if they occupy two neighboring sites this changes the pattern of virtual excitations, as is schematically illustrated in Fig.~\ref{fig:latticeExcitons}. In the Appendix, the energies of these various different configurations are calculated using second-order perturbation theory and assuming a hopping energy $t_{ab}$ between different orbitals on neighboring sites. The nearest-neighbor interaction energy $\Gamma_{nn}$ is defined as the energy gained by two holes when they benefit from these virtual excitations sitting on two-neighboring sites, rather than at an infinite distance from each other [see Eq.~(\ref{eq:nneighbors})].

It is convenient to introduce the dimensionless quantities
\begin{equation}
    u = \frac{U}{\Omega_{ab}};\quad
    j = \frac{J}{\Omega_{ab}};\quad
    \delta_{j} = \frac{\Delta_{J}}{\Omega_{ab}};\quad
    \gamma_{nn} = \Gamma_{nn}\frac{\Omega_{ab}}{t_{ab}^2}.
\end{equation}
We further note that
\begin{equation}
    \delta_{j} = \sqrt{1+j^2}-1\quad\text{and}\quad
    \alpha^2 = \frac{1}{2} + \frac{1}{2\sqrt{1+j^2}}.
\end{equation}
With these definitions, we express the nearest-neighbor interaction, Eq.~(\ref{eq:nneighbors}), as a dimensionless quantity
\begin{equation}
    \gamma_{nn} = \frac{\alpha^2}{1+\delta_j-j}+\frac{3\alpha^2}{1+\delta_j-3j}-\frac{4\alpha^4}{1+2\delta_j+u-5j}.
    \label{eq:uj0}
\end{equation}

\section{Discussion}

\begin{figure}[tb]
    \centering
    \includegraphics[width=0.9\columnwidth]{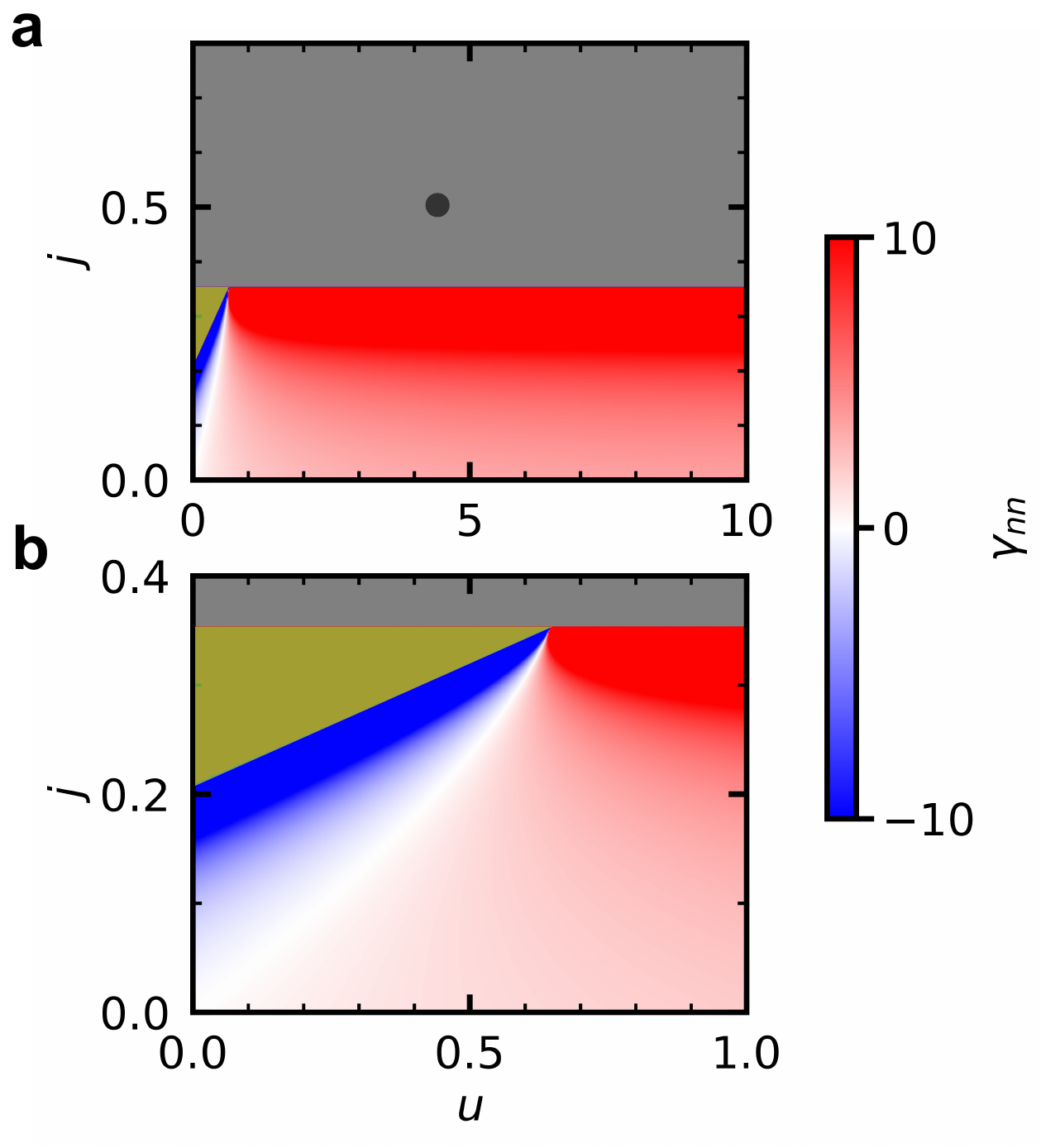}
    \caption{{\textbf{a}} Exciton mediated hole-hole interaction $\gamma_{nn}$ as function of the dimensionless $u$ and $j$ using Eq.~(\ref{eq:uj0}). {\textbf{b}} Zoom of panel {\textbf{a}}. The black dot in the top panel shows a realistic combination of parameters for Bi2212. The gray and moss green areas indicate where perturbation theory cannot be applied.}
    \label{fig:el_ex_interaction}
\end{figure}

A color map of $\gamma_{nn}$ in the $(u,j)$ plane is shown in Fig.~\ref{fig:el_ex_interaction}. Blue (red) indicates zones of attractive (repulsive) interaction. The gray color indicates the region where the ground state of a doped hole becomes unstable toward a different orbital configuration. This happens for $j\geqslant (1+\delta_j)/3$. For the purpose of our discussion we will only need to consider $j<(1+\delta_j)/3$. In this parameter range, $\delta_j\ll 1$ and $\alpha^2\approx 1$.

The experimental values of the $dd$ exciton peak observed by RIXS indicate that $\Omega_{ab}=2$~eV.
Following Ghijsen \textit{et al.} \cite{ghijsen1990}, we find for the cuprates $U=8.8$~eV and $J=1$~eV, so that $u=4.4$ and $j=0.5$. As pointed out above, this high value of $j$ is outside the range of validity of Eq.~(\ref{eq:uj0}). This should not come as a surprise: it is a manifestation of the second Hund's rule that --- in the absence of a crystal field --- the ground state is a high-spin state. For $j<1/3$ the crystal field is sufficiently strong to quench the high-spin state, but for the more realistic estimate $j=0.5$ the ground state has $S=1$. In the actual cuprate materials this does \emph{not} happen, because the energy of introducing a hole in O-$2p$ is lower than that of creating a $3d^8$ state \cite{zaanen1985}. In other words, if one considers realistic interaction parameters of the Cu-$3d$ electrons, it is necessary to take into account the O-$2p$ bands \cite{zhang1988,zaanen1985}. 

Meanwhile the question ``can $dd$ excitations mediate pairing ?'' has a non-trivial answer: There exists a set of parameters in the $(u,j)$ plane,
\begin{equation}
    u_c=j_c\frac{5-9j_c}{2-3j_c},
\end{equation}
for which $\gamma_{nn}=0$~\footnote{The expression for $u_c(j)$ is obtained by applying the
  condition $\gamma _{nn}(u_c,j)=0$ to Eq.~(\ref{eq:uj0})}. For $u>u_c(j)$ the interaction $\gamma_{nn}$ is repulsive; for $5j-1<u<u_c(j)$ it is attractive (always considering $j<1/3$).
For $u<5j-1$ (moss green area on the left of the color map) the state of the two doped holes on nearest-neighbor sites becomes unstable toward a different orbital configuration. As pointed out above, this model cannot be directly applied to the cuprates. On the other hand, as a matter of principle it \emph{is} of interest that such an attractive interaction is possible, and it may provide a method of tuning an attractive interaction in doped multi-band Mott insulators. Let us therefore take a step back and try to understand what exactly causes the attractive pairing channel.

From Eq.~(\ref{eq:uj0}) we see that in the parameter range $j<(1+u)/5$ and $u<u_c(j)$, the first and second terms of Eq.~(\ref{eq:uj0}) are both positive. When $j$ approaches $(1+u)/5$ from below, the denominator of the third term tends to zero and the attractive interaction potential diverges. Clearly the attractive interaction is caused by the third term. In the Appendix, we show that this term originates in virtual transitions that are unique to having two doped holes on nearest neighbor sites, i.e.\ $\underline{d}(R_1)^2\underline{d}(R_2)^2$. This opens a channel of virtual transitions to excited states $\underline{d}(R_1)^1\underline{d}(R_2)^3$. The energy cost of such a process is $\Omega_{ab}$ plus the effective on-site interaction energy $U_{\mathrm{eff}}=U-5J$. $U_{\mathrm{eff}}$ is \emph{negative} in the parameter range considered, $j<(1+u)/5$. For $j\rightarrow (1+u)/5$ the effective on-site interaction $U_{\mathrm{eff}} \rightarrow -\Omega_{ab}$. Consequently, the energies of the two states $\underline{d}(R_1)^2\underline{d}(R_2)^2$ and $\underline{d}(R_1)^1\underline{d}(R_2)^3$ approach the same value. One can therefore interpret this type of pairing interaction as a resonant process whereby two doped holes temporarily occupy the same site. It is not completely obvious if we should regard this as an exciton-mediated interaction; however, from a broader perspective, this interaction is effectively mediated by virtual local orbital fluctuations.

We close by making a speculation if this type of behavior could be realized in a real system, and if so, in which materials. From Fig.~\ref{fig:el_ex_interaction}, we see that in the region of attractive interaction we have $j\ll 1$, so that $u_c(j)\approx 5j/2$. The conditions $u<u_c(j)$ and $j<(1+u)/5$ then become
\begin{equation}
    5J -\Omega_{ab} < U \lesssim  5J/2.
    \label{eq:range}
\end{equation}
It is a well-known fact that the Hund's coupling $J$ is a relatively robust atomic parameter that is only weakly screened in a solid-state environment. In the $3d$ series, $J$ varies from 0.9~eV (Ti) to 1.5~eV (Cu), in the $4d$ series from 0.6~eV (Zr) to 1.0~eV (Ag), and in the $5d$ series from 0.7~eV (Hf) to 1.0~eV (Au) \cite{vandermarel1988}. The best targets are Nb$^{4+}$ or W$^{5+}$, both having a $d^1$ configuration. Taking $J=0.6$~eV and $\Omega_{ab}=2$~eV, $U$ should be about 1 to 1.5~eV. This is compatible with reported empirical values of the early $4d$ and $5d$ elements \cite{vandermarel1988}. However, since in this case the role of electrons and holes is reversed, the electron in a $d^1$ configuration occupies one of the $t_{2g}$ states. For an octahedral coordination the ground state is then 3-fold degenerate, making it Jahn-Teller active. This degeneracy can be lifted by the combination of spin-orbit coupling and a tetragonal crystal field. Electron-doped Sr$_2$NbO$_4$ or K$_2$WO$_4$ and other members of the Ruddleston-Popper series of these compounds could be candidate materials for observing the exciton mediated interactions of the type sketched above. To arrive at a quantitative prediction requires the implementation of spin-orbit coupling and a tetragonal crystal field.

\section{Conclusions}

We calculated the effective interaction between 2 holes doped in a 2-band Mott-insulator. We conclude from our study that the coupling to $dd$ excitations (orbital flips) mediates a nearest-neighbor interaction, the sign of which depends on the relative values of the on-site Coulomb repulsion and Hund's-rule exchange. Our analysis demonstrates that exciton-mediated interactions can in principle contribute positively to the pairing interaction, but depending on the parameters it may also lead to the opposite effect. To obtain a precise understanding it is necessary to include in the model the O-$2p$ states and the full set of Cu-$3d$ states. Better candidate materials for observing the type of interactions described here are members of the Ruddleston-Popper series of early transition-metal oxides such as Sr$_2$NbO$_4$ or K$_2$WO$_4$.

\begin{acknowledgements}
This work was supported by the Swiss National Science Foundation through Project No.\ 179157.
\end{acknowledgements}

\appendix*

\section{Details of the exciton mediated interaction}

\subsection{\boldmath The $\underline{d}^1$ states}
In the limit $U\rightarrow\infty$, the ground state of the undoped Mott insulator is the direct product of the eigenstates for each site, namely
\begin{equation}\begin{aligned}
    \ket{a_j} &= a^{\dagger}_{j\sigma} \ket{0}; & E_a &= \varepsilon_a\\
    \ket{b_j} &= b^{\dagger}_{j\sigma} \ket{0}; & E_b &= \varepsilon_a+\Omega_{ab}
    \label{eq:1hole}
\end{aligned}\end{equation}
where $\Omega_{ab}=\varepsilon_b-\varepsilon_a$.

\subsection{\boldmath The $\underline{d}^2$ states}
Hole doping causes a finite fraction of the sites to be occupied with 2 holes. 
For the case of 2 holes on a single site, the subspace with spin $S=0$ is spanned by the vectors
\begin{equation}\begin{aligned}
    \ket{aa;0} &= a^\dagger_{\uparrow} a^\dagger_{\downarrow} \ket{0}\\
    \ket{ab;0} &= \frac{1}{\sqrt{2}}(a^\dagger_{\uparrow} b^\dagger_{\downarrow}-a^\dagger_{\downarrow} b^\dagger_{\uparrow}) \ket{0}\\
    \ket{bb;0} &= b^\dagger_{\uparrow} b^\dagger_{\downarrow} \ket{0}.
\end{aligned}\end{equation}
On this basis, the Hamiltonian for $S=0$ is
\begin{equation}
    \hat{H}_{S=0}=\begin{pmatrix}
    2\varepsilon_a+U & 0 & J\\
    0 & \varepsilon_a+\varepsilon_b+U-J & 0\\
    J & 0 & 2\varepsilon_b+U
    \end{pmatrix}.
\end{equation}
Straightforward diagonalization gives the following set of eigenstates
\begin{equation}\begin{aligned}
    \ket{\tilde{aa};0} &= \big(\alpha a^\dagger_{\uparrow} a^\dagger_{\downarrow}-\beta b^\dagger_{\uparrow} b^\dagger_{\downarrow}\big)\ket{0}\\
    \ket{\tilde{bb};0} &= \big(\beta a^\dagger_{\uparrow} a^\dagger_{\downarrow}+\alpha b^\dagger_{\uparrow} b^\dagger_{\downarrow}\big)\ket{0}\\
    \ket{ab;0} &= \frac{1}{\sqrt{2}}\big(a^\dagger_{\uparrow} b^\dagger_{\downarrow}-a^\dagger_{\downarrow} b^\dagger_{\uparrow}\big) \ket{0},
\end{aligned}\end{equation}
where
\begin{equation}
    \alpha^2=\frac{1}{2}+\frac{\Omega_{ab}}{2\sqrt{\Omega_{ab}^2+J^2}},\quad
    \beta^2=\frac{1}{2}-\frac{\Omega_{ab}}{2\sqrt{\Omega_{ab}^2+J^2}}.
\end{equation}
The corresponding energies are
\begin{equation}\begin{aligned}
    E_{\tilde{aa};0} &= 2\varepsilon_a+U-\Delta_J\\
    E_{\tilde{bb};0} &= 2\varepsilon_a+2\Omega_{ab}+U+\Delta_J\\   
    E_{ab;0} & =2\varepsilon_a+\Omega_{ab}+U-J
\end{aligned}\end{equation}
with
\begin{equation}
    \Delta_J=\sqrt{\Omega_{ab}^2+J^2} - \Omega_{ab}.
\end{equation}
The basis vectors for the $S=1$ manifold are
\begin{equation}\begin{aligned}
    \ket{ab;1,1}  &= a^\dagger_{\uparrow} b^\dagger_{\uparrow} \ket{0}\\
    \ket{ab;1,0}  &= \frac{1}{\sqrt{2}}(a^\dagger_{\uparrow} b^\dagger_{\downarrow}+a^\dagger_{\downarrow} b^\dagger_{\uparrow}) \ket{0}\\
    \ket{ab;1,-1} &= a^\dagger_{\downarrow} b^\dagger_{\downarrow} \ket{0}.
\end{aligned}\end{equation}
On this basis, the Hamiltonian for $S=1$ is
\begin{equation}
    \hat{H}_{S=1}=
    \begin{pmatrix}
    E_{ab;1}  & 0 & 0\\
    0 & E_{ab;1}  & 0\\
    0 & 0 &  E_{ab;1} 
    \end{pmatrix}
\end{equation}
with
\begin{equation}
   E_{ab;1} = 2 \varepsilon_a + \Omega_{ab}+U-3J.
\end{equation}

\subsection{\boldmath The $\underline{d}^3$ states}
For the evaluation of virtual processes where 2 doped holes occupy nearest-neighbor sites, we will need the following $\underline{d}^3$ states:
\begin{equation}\begin{aligned}
    {\ket{aab;1/2,\sigma}} &= a^{\dagger}_{\uparrow}a^{\dagger}_{\downarrow}b^{\dagger}_{\sigma}\ket{0}\\
    E_{aab} &= 3\varepsilon_a + \Omega_{ab} + 3U-5J.
    \label{eq:3holes}
\end{aligned}\end{equation}

\subsection{Dressing of a doped hole by excitons on neighboring sites}
When a hole is introduced in the system, it can be ``dressed'' by creating virtual excitons on the neighboring sites. To estimate the energy saving by these processes, we consider a 2-site cluster where one hole is introduced in addition to the hole already present at each site:
\begin{equation}\begin{aligned}
    \ket{\psi_g^{1h}} &= \big(\alpha a^\dagger_{1\uparrow} a^\dagger_{1\downarrow}-\beta b^\dagger_{1\uparrow} b^\dagger_{1\downarrow}\big)  
    a^{\dagger}_{2\uparrow}\ket{0}\\
    E_g^{1h} &= E_{\tilde{aa};0} + E_a = 3\varepsilon_a+U-\Delta_{J},
\end{aligned}\end{equation}
where the indices $1,2$ refer to the different sites and the label $1h$ refers to 1 doped hole.
The relevant excited states are
\begin{equation}\begin{aligned}
    \ket{\psi_1^{1h}} &= \frac{1}{\sqrt{2}}\big( a^{\dagger}_{1\uparrow}a^{\dagger}_{2\uparrow}b^{\dagger}_{2\downarrow}
    - a^{\dagger}_{1\uparrow}a^{\dagger}_{2\downarrow}b^{\dagger}_{2\uparrow}\big)\ket{0}\\
    E_{1}^{1h} &= E_{ab;0} + E_a = E_g^{1h}+\Delta_{J}+\Omega_{ab}-J\\
    \ket{\psi_2^{1h}} &= \frac{1}{\sqrt{2}}\big( a^{\dagger}_{1\uparrow}a^{\dagger}_{2\uparrow}b^{\dagger}_{2\downarrow}
    + a^{\dagger}_{1\uparrow}a^{\dagger}_{2\downarrow}b^{\dagger}_{2\uparrow}\big)\ket{0}\\
    E_{2}^{1h} &= E_{ab;1} + E_a = E_g^{1h}+\Delta_{J}+\Omega_{ab}-3J\\
    \ket{\psi_3^{1h}} &= a^{\dagger}_{1\downarrow}a^{\dagger}_{2\uparrow}b^{\dagger}_{2\uparrow}\ket{0}\\
    E_{3}^{1h} &= E_{ab;1} + E_a = E_g^{1h}+\Delta_{J}+\Omega_{ab}-3J.
\end{aligned}\end{equation}
The intersite hopping term $\hat{H}_t$ connecting different orbitals $a$ and $b$ is
\begin{equation}
    \hat{H}_t = t_{ab} \sum_{\sigma} \left(a^{\dagger}_{1\sigma}b^{}_{2\sigma}+b^{\dagger}_{1\sigma}a^{}_{2\sigma} + \mathrm{h.c.}\right).
\end{equation}
We will need below the matrix elements of the hopping term $\hat{H}_t$, for which we obtain
\begin{align}
    \nonumber
    |\bra{\psi_g^{1h}} \hat{H}_t \ket{\psi_1^{1h}}|^2 
    &= |\bra{\psi_g^{1h}} \hat{H}_t \ket{\psi_2^{1h}} |^2  = \frac{1}{2} \alpha^2 t_{ab}^2\\
    |\bra{\psi_g^{1h}} \hat{H}_t \ket{\psi_3^{1h}}|^2 
    &= \alpha^2 t_{ab}^2.
\end{align}
With the help of these expressions, we calculate the energy lowering of a site with a single doped hole due to virtual hopping to a different \footnote{We do not consider contributions from virtual hopping to the same orbital on a neighboring site, since --- important as they are --- these processes are accounted for by the one-band Hubbard model.} orbital on a neighboring site
\begin{equation}
    \tilde{E}_g^{1h} = E_g^{1h} - \frac{\alpha^2t_{ab}^2/2}{\Omega_{ab}-J+\Delta_{J}} - \frac{3\alpha^2t_{ab}^2/2}{\Omega_{ab}-3J+\Delta_{J}}.
     \label{eq:onehole}
\end{equation}

\subsection{Two holes at nearest-neighbor sites}
If we introduce two doped holes on nearest-neighbor sites, as described by the state vector
\begin{align}
    \nonumber
    \ket{\psi_g^{2h}} &= 
    \big(\alpha a^\dagger_{1\uparrow} a^\dagger_{1\downarrow}-\beta b^\dagger_{1\uparrow} b^\dagger_{1\downarrow}\big)
    \big(\alpha a^\dagger_{2\uparrow} a^\dagger_{2\downarrow}-\beta b^\dagger_{2\uparrow} b^\dagger_{2\downarrow}\big)\ket{0}\\
    E_g^{2h} &= 2E_{\tilde{aa};0} = 4\varepsilon_a+2U-2\Delta_{J},
\end{align}
the virtual excitations whereby a hole hops from one of these sites to the other are modified, because the intermediate state on one of the two sites now has $2$ doped holes in addition to the hole already present in the undoped state (\textit{i.e.}\ $3$ holes in total). Specifically these virtual states are
\begin{equation}\begin{aligned}
    \ket{\psi_1^{2h}} &= a^{\dagger}_{1\uparrow}a^{\dagger}_{2\uparrow}a^{\dagger}_{2\downarrow}b^{\dagger}_{2\downarrow}\ket{0}\\
    \ket{\psi_2^{2h}} &= a^{\dagger}_{1\downarrow}a^{\dagger}_{2\uparrow}a^{\dagger}_{2\downarrow}b^{\dagger}_{2\uparrow}\ket{0}\\
    \ket{\psi_3^{2h}} &= a^{\dagger}_{1\uparrow}a^{\dagger}_{1\downarrow}a^{\dagger}_{2\uparrow}b^{\dagger}_{1\downarrow}\ket{0}\\
    \ket{\psi_4^{2h}} &= a^{\dagger}_{1\uparrow}a^{\dagger}_{1\downarrow}a^{\dagger}_{2\downarrow}b^{\dagger}_{1\uparrow}\ket{0}
   \label{eq:twosite} 
\end{aligned}\end{equation}
with energy
\begin{align}
    E_{j}^{2h} &= E_{aab} + E_a\\
    \nonumber
    &= E_g^{2h}+2\Delta_{J}+\Omega_{ab} + U - 5J 
    \quad (1\leqslant j \leqslant 4).
    \label{eq:twosite-energies} 
\end{align}
As before, we have to compute the matrix elements of the hopping term $\hat{H}_t$, with the result
\begin{equation}
    |\bra{\psi_g^{2h}}\hat{H}_t\ket{\psi_j^{2h}}|^2 = \alpha^4t_{ab}^2 \qquad (1\leqslant j \leqslant 4).
    \label{eq:intratwosite}
\end{equation}
We calculate the correction to the ground-state energy due to virtual hopping between the two sites sketched in Fig.~\ref{fig:latticeExcitons}\textbf{b} and subtract the contribution from the virtual hopping processes that they replace [see Eq.~(\ref{eq:onehole})]. This way we obtain for the energy of bringing two holes from an infinite distance to a nearest-neighbor distance
\begin{multline}
    \Gamma_{nn} = \frac{\alpha^2t_{ab}^2}{\Omega_{ab}-J+\Delta_{J}}+\frac{3\alpha^2t_{ab}^2}{\Omega_{ab}-3J+\Delta_{J}}\\
    -\frac{4\alpha^4t_{ab}^2}{\Omega_{ab}+U-5J+2\Delta_{J}}.
    \label{eq:nneighbors}    
\end{multline}

%
%
\end{document}